\documentstyle[amsfonts,epsf,aps]{revtex}
\newcommand{\ee}{\end{equation}}
\newcommand{\be}{\begin{equation}}

\begin{document}

\baselineskip 2\baselineskip
\title{Deterministic  motion of the controversial piston 
in the thermodynamic limit\vskip 3mm}
\author{ Christian Gruber and S\'everine Pache}

\address{\vskip 3mm Institut de Physique Th\'{e}orique, Ecole Polytechnique
F\'ed\'erale de Lausanne, CH-1015 Lausanne, Switzerland}

\author{\vskip 1mm Annick Lesne}

  \address{\vskip 5mm
Laboratoire de Physique Th\'eorique 
des Liquides, 
Universit\'e Pierre et Marie Curie,  
Case courrier 121,\\\vskip 3mm4 Place Jussieu, 75252
 Paris Cedex 05,
France\vskip 6mm}
\date{\today}
\maketitle
\vskip 3mm
\begin{abstract}
{\small
\baselineskip 2\baselineskip
We consider the evolution of a system composed of $N$ non-interacting point 
particles of mass $m$ in a cylindrical container divided into 
two regions by a movable adiabatic wall (the adiabatic piston).
 We study the thermodynamic limit for the piston where the area $A$ 
of the  cross-section, the mass $M$ of the piston, and  the number
 $N$ of particles go to infinity keeping $A/M$ and $N/M$ fixed. The 
length   of the container is a fixed parameter which can be either
 finite or infinite. In this thermodynamic limit we show 
that the  motion of the piston is deterministic and the evolution 
is adiabatic. Moreover if the length of the container is infinite,
 we show that the piston evolves toward a stationary state with velocity
 approximately
proportional
to the pressure difference. If the length of the container is finite, 
introducing a simplifying assumption we show that the system evolves
 with either weak or strong damping toward a well-defined state of
 mechanical equilibrium where the pressures are the same, but
 the temperatures different. Numerical simulations are presented 
to illustrate possible evolutions and to check the validity of the assumption.

}
\end{abstract}

{\it  {\it \fontshape{it} Key Words:}
Liouville equation, adiabatic, piston, equilibrium, damping.}

\section*{  1. Introduction}

The adiabatic piston problem is a well-known controversial example of
thermodynamics where the two principles of thermostatics (conservation 
of energy  and maximum of entropy) are not sufficient to obtain the final state
 to which an isolated system will evolve. This can be understood since
 the final state will depend upon the values of the ``friction coefficients'',
which however do not appear in the  entropy function. Similarly, taking a microscopical
approach, i.e. statistical mechanics, this implies that one can not find the 
(thermodynamical) equilibrium state by phase space arguments.

Let us recall the problem. An  isolated, rigid cylinder is filled with 
two 
ideal
gases separated by a movable adiabatic rigid wall (the piston). Initially
the piston is fixed by a brake at some position $X_0$ and 
the two gases are in thermal equilibrium characterized by their respective
pressures $p^{\pm}$ and 
 temperatures $T^{\pm}$
(or equivalently their energies $E^{\pm}$). At a certain
time, the brake is released and the question is to find the final
 equilibrium state assuming that  no friction is involved in
 the microscopical dynamics (see [1-3] and references therein
 for different discussions). Experimentally the ``adiabatic piston" has
 been used already before $1940$ \cite{refex} to measure  the ratio of the specific heat of gases $\gamma =c_p/c_v$.

Recently, the problem was investigated using the following very simple
microscopic model [5-9], which is a variation of the 
 model already introduced in \cite{lebowitz59}.
The  system consists  of $N$ non-interacting point particles of mass $m$ in an
adiabatic, rigid cylinder of length $2L$ and cross-section $A$.
The cylinder is divided in two compartments,
containing respectively $N^-$ and $N^+$ particles,
 by a movable wall of mass $M\gg m$,
with no internal degrees of freedom (i.e. an adiabatic piston).
 This piston is constrained to
move without friction along the $x$-axis. The particles make purely elastic collisions
on the boundaries of the cylinder and on the piston,
i.e. if    $v$ and 
$V$  denote the $x$-component of the velocities 
of a particle and the piston before a
collision, then under collision on the piston:
\be\label{vprime}
v\rightarrow v^{\prime}=2V-v+\alpha(v-V)\hskip 10mm
V\rightarrow V^{\prime}=V+\alpha(v-V)\ee
where:
\be
\alpha = {2m\over M+m}
\ee
Similarly, for a collision of a particle with the boundaries 
at $x=\pm L$, we have:
\be\label{vprime2}
v\rightarrow v^{\prime}=-v\ee
Since there is no coupling with the transverse degrees of freedom, one can assume that
all
probability distributions are independent of the  tranverse coordinates.
We are thus led to a formally one-dimensional problem 
(except for normalizations) and therefore
in the model we shall assume that all the particles have 
velocities parallel to
the $x$-axis.

Since  in physical situations $m\ll M$, this model was investigated
in \cite{GP} in the limit where $\epsilon=\sqrt{m/M}\ll 1$
with $M$ fixed. It was shown that if initially the two gases 
have the same pressure but different temperatures, i.e. $p^+=p^-$
but $T^+\neq T^-$, then for the infinite cylinder ($L=\infty$)
and in the limit $\epsilon\rightarrow 0$ ($M$ fixed, finite), the stationary
solution of Boltzmann equation describes an equilibrium state 
and the velocity distribution for the piston is Maxwellian, with temperature
$T_p=\sqrt{T^-T^+}$, i.e. the piston is an adiabatic wall.
However, to first order in $\epsilon$, the stationary solution of
Boltzmann equation is no longer Maxwellian and describes a non-equilibrium
state where the piston moves with constant average velocity
$\bar{V}=\sqrt{\pi mk_B/8}(\sqrt{T^+}-\sqrt{T^-})/M$
towards the high temperature domain, although the pressures are equal.
In \cite{pache}, it was shown that for $L=\infty$, the evolution towards the
stationary state is described to order zero in $\epsilon$
by a Fokker-Planck equation, which can then be used to study the evolution
to higher order in $\epsilon$.
The case of a finite cylinder was investigated in \cite{frache}
by qualitative arguments and numerical simulations. It was shown that 
the evolution take place in two stages with very different time scales.
In the first stage, the evolution is adiabatic and proceeds rather rapidly, 
with or without oscillations, to a state of ``mechanical equilibrium''
 where the pressures are equal but the temperatures different.
In the second stage the evolution takes place on a time scale several orders
of magnitude  larger (if $\epsilon\ll 1$);
in this second stage, the piston drifts very slowly towards
the high temperature domain, the pressures of both gases remain approximately constant,
and
the temperatures vary very slowly to reach a final equilibrium
state where densities, pressures and temperatures of the two gases are the same.
It was thus concluded in \cite{frache} that a wall which is adiabatic when fixed, becomes
heat-conducting under the stochastic motion. However
for real systems, the time involved to reach the thermal equilibrium
will be several million times the age of universe and thus for all practical
purposes the piston is in fact an adiabatic wall.

On the other hand, for systems with $\epsilon\approx 1$, or $\epsilon=1$
[11-14],
the system evolves directly toward thermal equilibrium and can never
be considered as an adiabatic wall.

Another limit has been recently considered.
In \cite{LPS}, the authors have studied the case where the mass
$m$  of the particles is fixed, but $L$ tends to infinity
together with $M\sim L^2$, $N\sim L^3$ and the time is scaled with
$t=\tau L$. They have shown that in this limit, the motion
of the piston and the one-particle distribution of the gas satisfy 
autonomous coupled equations.

Following a suggestion of J.L. Lebowitz \cite{JLL}, we shall analyse in the 
following still another limiting procedure. We consider the thermodynamic
 limit for the piston where $m$ and $L$ are fixed, but the area $A$
of the cross-section of the cylinder tends to infinity while
\be
\gamma={2mA\over M+m}
\hspace{10mm}{\rm and}\hspace{10mm} R^{\pm}=m\;{N^{\pm}\over M}
\ee
are kept constant (``$-$'' refers to  the left  and ``$+$''
 to the right  of the piston).



The object of this article is to show that in the thermodynamic limit for the piston, 
the motion of the piston is ``adiabatic'' in the sense that it is deterministic,
i.e. $<V^n>_t=<V>^n_t$, no heat transfer is involved, the entropy of both gases 
increases, there is factorization of the joint distribution for one particle and the piston, and the system evolves toward a state
of mechanical equilibrium, which is not a state of thermal equilibrium.
Furthermore numerical simulations indicate that the evolution
is such that the energy  of the gas increases under compression
(because work is done on the gas); moreover the simulations also
indicate that the evolution depends strongly on $R^{\pm}$ for small
values ($R^{\pm}<10$) but tends to be independent of $R^{\pm}$
for larger values (see Figure $2$ and $3$).

\vskip 1mm
\noindent
In section $2$ we derive coupled equations for the one-particle
 velocity distributions. The thermodynamic limit for the piston
 is investigated in section $3$. In section $4$ we discuss the case
 where the length of the container is infinite; then the case of
 a finite container is considered in section $5$. We present
 numerical simulations in section $6$, and finally the conclusions in the last section.

\section*{ 2. Coupled equations for the one-particle velocity distributions}\label{sec2}

\subsection{Initial condition}

We label $(-)$ those properties associated with the particles 
in the left compartment  and $(+)$ those associated with the right compartment.
 The 
number $N^-$ and $N^+$ of particles in each compartment is fixed.
For the sake of clarity, when necessary we denote
$(x_i,
v_i)$, $i=1\ldots N^-$,  the position and velocity of the  particles in the
 left compartment
(``left particles"), by $(y_j,
w_j)$, $j=1\ldots N^+$,  the position and velocity of the  right particles,
and by $(X,V)$ the position and velocity  of the piston.
Taking the boundaries of the cylinder at $x=\pm L$, we thus have:
\be
-L\leq x_i\leq X\leq y_j\leq L
\ee
In this section,  $L$, $N^{\pm}$ and $M$ are finite.
Initially, the piston is fixed at ($X=X_0$, $V=0$) and the particles on both sides
 are in thermal equilibrium at respective  temperatures 
$T^-$ and $T^+$, i.e. the probability distribution $f$ 
in the whole phase space is
given by :
\vskip 3mm
$
f(x_1,v_1;\ldots; x_{N^-}, v_{N^-}; X, V; y_1,w_1;\ldots; y_{N^+},w_{N^+};t=0)=
$
\be
={(N^-)!\over [A(L+X_0)]^{N^-}}\;
{(N^+)!\over [A(L-X_0)]^{N^+}}\;\prod_{i=1}^{N^-}f^-(v_i)\theta(X_0-x_i)
\theta(x_i+L)\;
\prod_{j=1}^{N^+}f^+(w_j)\theta(y_j-X_0)
\theta(L-y_j)\delta(X-X_0)\delta(V)
\ee
where $\theta$ denotes the Heaviside step function.
The probability distributions
 $f^{\pm}(v)$ are functions of $v^2$ with:
\be
\int_{-\infty}^{\infty}f^{\pm}(v)dv=1, 
\hspace{10mm}
2m\int_{-\infty}^0v^2f^{+}(v)dv=k_BT^{+},
\hspace{10mm}2m\int_{0}^{\infty}v^2f^{-}(v)dv=k_BT^{-}\ee
For example, one could take for  $f^{\pm}(v)$ Maxwellian distributions with
temperatures $T^{\pm}$, or functions which are non-zero only for $|v|\in[v_{min},
v_{max}]$.
For $t\geq 0$, the piston moves freely and the problem is to study its evolution.

\subsection{Liouville equation}
The distribution 
function  $f=f(\{x_i,v_i\};X,V;\{y_j,w_j\};t)$ at time $t$
is a measure over the whole phase space, which is  symmetric in the
$i$-coordinates (resp. in the $j$-coordinates), 
with  the following normalization which 
 reflects the underlying dimension $d=3$ of
the system:
\be
\int
f(x_1,v_1;\ldots; x_{N^-}, v_{N^-}; X, V; y_1,w_1;\ldots; y_{N^+},w_{N^+};t)\;
dXdV\prod_{i=1}^{N^-}dx_1dv_1\prod_{j=1}^{N^+}dy_jdw_j=
{(N^-)!\;(N^+)!\over A^{N^-+N^+}}
\ee
(with $A$ the cross-section of the cylinder).
Its evolution in time 
is given by the Liouville equation: 
\be\label{Liouville}
\left({\partial\over \partial t}
+\sum_{i=1}^{N^-}v_i{\partial\over \partial x_i}+
V{\partial\over \partial X}+
\sum_{j=1}^{N^+}w_j {\partial\over \partial y_j}\right)f={\cal I}
\ee
together with the boundary conditions:
\be\label{bc1}
f(\ldots)_{|x_i=-L, v_i}=f(\ldots)_{|x_i=-L, -v_i}
\ee
\be\label{bc2}
f(\ldots)_{|y_j=L, w_j}=f(\ldots)_{|y_j=L, -w_j}
\ee
The right-hand side
${\cal I}$ of Eq. (\ref{Liouville})
takes into account the  elastic
collisions of the  particles with the wall at $\pm L$ and with the piston.
It decomposes into four terms.
\vskip 1mm
\noindent
\underline{\it 1 - Elastic collisions of left particles with the wall at $x=-L$:}
\vskip 3mm $
{\cal I}_1=\sum_{i=1}^{N^-}\delta(x_i+L)v_i
[\theta(v_i)f(x_1,v_1;\ldots;x_i,-v_i;\ldots; x_{N^-},
 v_{N^-}; X, V; y_1,w_1;\ldots;
 y_{N^+},w_{N^+};t)$
\be +\theta(-v_i)f(x_1,v_1;\ldots;x_i,v_i;\ldots; x_{N^-}, v_{N^-};
 X, V; y_1,w_1;\ldots;
 y_{N^+},w_{N^+};t)]
\ee
 Using the boundary condition (\ref{bc1}), this contribution is simply:
\be
{\cal I}_1=\sum_{i=1}^{N^-}\delta(x_i+L)v_if
\ee

\vskip 1mm
\noindent
\underline{\it 2 - Elastic collisions of right particles with the wall at $x=L$:}
\vskip 3mm ${\cal I}_2=-\sum_{j=1}^{N^+}\delta(y_j-L)w_j
[\theta(-w_j)f(x_1,v_1;\ldots; x_{N^-},
 v_{N^-}; X, V; y_1,w_1;\ldots;y_j,-w_j;\ldots;
 y_{N^+},w_{N^+};t)$
\be+\theta(w_j)f(x_1,v_1;\ldots;
 x_{N^-}, v_{N^-}; X, V; y_1,w_1;\ldots;y_j,w_j;\ldots;
 y_{N^+},w_{N^+};t)]
\ee
and using the boundary condition (\ref{bc2}), we have:
\be
{\cal I}_2=-\sum_{j=1}^{N^+}\delta(y_j-L)w_jf
\ee
\vskip 1mm
\noindent
\underline{\it 3 - Elastic collisions of left particles with the piston at $x=X$:}
\vskip 3mm $
{\cal I}_3=\sum_{i=1}^{N^-}\delta(x_i-X)(V-v_i)
[\theta(V-v_i)f(x_1,v_1;\ldots;x_i,v_i^{\prime};\ldots; x_{N^-},
 v_{N^-}; X, V^{\prime}; y_1,w_1;\ldots;
 y_{N^+},w_{N^+};t)$
\be +\theta(v_i-V)f(x_1,v_1;\ldots;x_i,v_i;\ldots; x_{N^-}, v_{N^-}; X, V;
 y_1,w_1;\ldots;
 y_{N^+},w_{N^+};t)]
\ee
where $(v_i^{\prime}, V^{\prime})$ are given
 by Eqs. (\ref{vprime}, $2$).
\vskip 3mm
\noindent
\underline{\it 4- Elastic collisions of right particles with the piston at $y=X$:}
\vskip 3mm $
{\cal I}_4=-\sum_{j=1}^{N^+}\delta(y_j-X)(V-w_j)
[\theta(w_j-V)f(x_1,v_1;\ldots; x_{N^-},
 v_{N^-}; X, V^{\prime}; y_1,w_1;\ldots;y_j,w_j^{\prime};\ldots;
 y_{N^+},w_{N^+};t)$
\be+\theta(V-w_j)f(x_1,v_1;\ldots;
 x_{N^-}, v_{N^-}; X, V; y_1,w_1;\ldots;y_j,w_j;\ldots;
 y_{N^+},w_{N^+};t)]
\ee
The correlation functions for $k$ left particles,
$l$ right particles (and possibly the piston) are defined by integration of the
distribution $f$ over $(N^--k)$ left particles, 
$(N^+-l)$ right particles, and the piston coordinates (possibly not),
multiplied by the factor:  
\be
{A^{N^--k}\over(N^--k)!}\;{A^{N^+-l}\over (N^+-l)!}
\ee
The evolution for the correlation functions is then obtained from Liouville equation
by integration over the corresponding coordinates.

\subsection{Equation for $\rho^-(x,v;t)$}

We denote by  $\rho^-(x,v;t)$ the single-particle
 distribution in the left compartment,
obtained  by integrating over all variables except $x_1=x$ and $v_1=v$,
with the normalization:
\be
\int\rho^-(x,v;t)dxdv={N^-\over A}
\ee
Its evolution involves $\rho_{1,P}$  the correlation function 
for one left particle and the piston, with:
\begin{eqnarray}
& &\int \rho_{1,P}(x,v;X,V;t)dXdV=\rho^-(x,v;t)\label{16}\\
& &\int \rho_{1,P}(x,v;X,V;t)dxdv={N^-\over A}\;\Psi(X,V;t)\label{17}
\end{eqnarray}
where $\Psi(X,V;t)$ is the normalized probability distribution for the piston,
obtained by integrating $f$ over all variables except $(X,V)$.
Integrating Liouville equation (\ref{Liouville}) over all variables
 except $x_1=x$
and $v_1=v$, together with the boundary conditions
(\ref{bc1}) yields:
\vskip 3mm $
(\partial_t+v\partial_x)\rho^-(x,v;t)=\delta (x+L/2)v\rho^-(x,v;t)$
\be\label{evolrho-}
+\int_{-\infty}^{\infty}(V-v)[\theta(V-v)
\rho_{1,P}(x,v^{\prime};x,V^{\prime};t)+\theta(v-V)\rho_{1,P}(x,v;x,V;t)]dV
\ee
with the initial condition:
\be
\rho^-(x,v;t=0)=f^-(v)\;{N^-\over A(L+X_0)}\;
\theta(X_0-x)\theta(x+L)
\ee

\subsection{Equation for $\rho^+(y,w;t)$} 
Similarly, we introduce  $\rho^+(y,w;t)$ the single-particle distribution in the right
compartment and $\rho_{P,1}$  the correlation function 
for one right-particle and the piston:
\begin{eqnarray}
& & \int \rho^+(y,w;t)dydw={N^+\over A}\\
& & \int \rho_{P,1}(X,V;y,w;t)dXdV=\rho^+(y,w;t)\\
& & \int \rho_{P,1}(X,V;y,w;t)dydw={N^+\over A}\;\Psi(X,V;t)
\end{eqnarray}
Integrating Liouville equation (\ref{Liouville}) together with 
the boundary condition (\ref{bc2}) yields:
\vskip 3mm $
(\partial_t+w\partial_y)\rho^+(y,w;t)=-\delta (y-L)w\rho^+(y,w;t)$
\be\label{evolrho+}
-\int_{-\infty}^{\infty}(V-w)[\theta(w-V)
\rho_{P,1}(y,V^{\prime};y,w^{\prime};t)+\theta(V-w)\rho_{P,1}(y,V;y,w;t)]dV
\ee
together with the initial condition:
\be
\rho^+(y,w;t=0)=f^+(w)\;{N^+\over A(L-X_0)}\;
\theta(y-X_0)\theta(L-y)
\ee

\subsection{Equation for $\Psi(X,V;t)$ and $\Phi(V;t)$} 
Integrating Liouville equation (\ref{Liouville})
 over all variables except $(X,V)$, and assuming
$\Psi(X=\pm L, V;t)=0$, yields:
\vskip 3mm \begin{eqnarray}
(\partial_t+V\partial_X)\Psi(X,V;t)=&A&\;\int_{-\infty}^{\infty}
(V-v)[\theta(V-v)\rho_{1,P}(X,v^{\prime};X,V^{\prime};t)
+\theta(v-V)\rho_{1,P}(X,v;X,V;t)]dv \nonumber \\
-&A&\;\int_{-\infty}^{\infty}
(V-w)[\theta(w-V)
\rho_{P,1}(X,V^{\prime};X,w^{\prime};t)+\theta(V-w)\rho_{P,1}(X,V;X,w;t)]dw\label{Psi}
\end{eqnarray}
together   with the initial condition:
\be
\Psi(X,V;t=0)=\delta(X-X_0)\delta(V)
\ee
Finally integrating  (\ref{Psi}) over $X$
leads to the evolution equation of the distribution function 
$\Phi(V;t)$ for the velocity  of the piston:
\begin{eqnarray}
\partial_t\Phi(V;t)=
&A&\;\int_{-\infty}^{\infty}
(V-v)[\theta(V-v)\rho_{surf}^-(v^{\prime};V^{\prime};t)
+\theta(v-V)\rho_{surf}^-(v;V;t)]dv \nonumber\\
\label{partialphi1}
-&A&\;\int_{-\infty}^{\infty}
(V-w)[\theta(w-V)
\rho_{surf}^+(w^{\prime};V^{\prime};t)+\theta(V-w)\rho_{surf}^+(w;V;t)]dw
\end{eqnarray}
where:
\begin{eqnarray}
\rho_{surf}^-(v;V;t)&=&\int_{-\infty}^{\infty} \rho_{1,P}(X,v;X,V;t)dX\label{29}\\
\rho_{surf}^+(w;V;t)&=&\int_{-\infty}^{\infty} \rho_{P,1}(X,V;X,w;t)dX\label{30}
\end{eqnarray}
represent the joint distribution for the piston velocity $V$ and the particle velocity $v$
on the left (resp. right) surface of the piston.
Let us note that
\be \rho^{\pm}_{surf}(t)=\int_{-\infty}^{\infty}\int_{-\infty}^{\infty}
\rho_{surf}^{\pm}(v;V;t)dvdV\ee
represents the density of particles on the left (resp. right) surface of the piston.

\vskip 5mm

Let us transform  equation (\ref{partialphi1}) by considering its action 
$<\Phi,g>(t)=\int_{-\infty}^{\infty}\Phi(V;t)g(V)dV$
on a test-function $g(V)$.
We have by definition:
\be\label{32}
{d\over dt}<\Phi,g>=\int_{-\infty}^{\infty}\partial_t\Phi(V;t)g(V)dV
\ee
Introducing the  expression  (\ref{partialphi1}) of $\partial_t\Phi(V;t)$ 
in Eq. (\ref{32}) and
making the change of integration variables
$(v,V)\rightarrow (v^{\prime},V^{\prime})$ in the term involving
$\rho_{surf}^-(v^{\prime};V^{\prime};t)$ and, respectively,
the change of integration variables
$(V,w)\rightarrow (V^{\prime},w^{\prime})$
in the term involving $\rho_{surf}^+(V^{\prime};w^{\prime};t)$),
we obtain:
\begin{eqnarray}
{d\over dt}<\Phi,g>=
&A&\int_{-\infty}^{\infty}
dVdv(v-V)\theta(v-V)\rho_{surf}^-(v;V;t)[g(V+\alpha(v-V))-g(V)]
\nonumber\\ 
-& A&\int_{-\infty}^{\infty} dVdw (w-V)\theta(V-w) \rho_{surf}^+(w;V;t) [g(V+\alpha(w-V))-g(V)]
\end{eqnarray}
We then  expand $g(V+\alpha(v-V))$ in powers of $\alpha$:
\be g(V+\alpha(v-V))-g(V)=\alpha\sum_{k=0}^{\infty}{\alpha^k(v-V)^{k+1}\over (k+1)! }
g^{(k+1)}(V)\ee
and  transform the terms as follows:
\vskip 3mm $
\alpha A\int_{-\infty}^{\infty}
dVdv(v-V)\theta(v-V)\rho_{surf}^-(v;V;t){(v-V)^{k+1} }
g^{(k+1)}(V)=$\enlargethispage*{1\baselineskip}
\be
(-1)^{k+1}\,\gamma\left\langle 
{\partial^{k+1}\over \partial V^{k+1}}\Big[\int_V^{\infty}
(v-V)^{k+2}\rho_{surf}^-(v;V;t)dv\Big],g \right\rangle
\ee
with 
\be\gamma=\alpha A={2mA\over M+m}
\ee
In conclusion, 
the equation for the distribution $\Phi(V;t)$ is
\be\label{partialphi2}
\partial_t\Phi(V;t)=-\,
\gamma\;\sum_{k=0}^{\infty}{(-1)^{k}\alpha^k\over (k+1)!} 
{\partial^{k+1}\over \partial V^{k+1}}
\left[\int_V^{\infty}
(v-V)^{k+2}\rho_{surf}^-(v;V;t)dv -\int_{-\infty}^V
(v-V)^{k+2}\rho_{surf}^+(v;V;t)dv
\right]\ee

\section*{3. Thermodynamic limit for the piston}

The thermodynamic limit for the piston is defined by 
$A\rightarrow\infty$, $M\rightarrow\infty$,
$N^{\pm}\rightarrow\infty$ with $\gamma=2mA/(M+m)$
and $R^{\pm}=N^{\pm}m/M$ fixed.
In this limit,  $\alpha=0$
and the  equations for collisions (\ref{vprime}), ($2$) are simply:
\be
v^{\prime}=2V-v\hspace{10mm}V^{\prime}=V
\ee
Similarly, the boundary conditions for the one-particle correlation
functions write:
\begin{eqnarray}
\rho^-(X,v;X,V;t)&=&\rho^-(X,(2V-v);X,V;t)\\
\rho^+(X,V;X,w;t)&=&\rho^+(X,V;X,(2V-w);t)
\end{eqnarray}
Assuming that we can permute $\alpha\rightarrow 0$ with the sum over $k$ 
in Eq. (\ref{partialphi2}), which we expect to be verified for initial conditions such that $<V>_t$ is not too large for all $t$, the equation for the evolution of $\Phi(V;t)$
reduces to:
\be
\partial_t\Phi(V;t)=-\,\gamma\;{\partial\over \partial V}
\left(\int_V^{\infty}
(v-V)^{2}\rho_{surf}^-(v;V;t)dv -\int_{-\infty}^V
(v-V)^{2}\rho_{surf}^+(v;V;t)dv\right)
\ee
With the initial condition  we have considered, it is natural to 
expect  that it is
possible to write:
\be\label{ac}
\rho_{surf}^{\pm}(v;V;t)=a^{\pm}(v;V;t)\Phi(V;t)\ee
where $a^{\pm}(v;V;t)$ are 
non-negative continuous functions of $V$, while $\Phi(V;t)$ is a
distribution in $V$ for all $t$.
As we shall see, the assumption (\ref{ac}) will be justified {\it a fortiori}
after the discussion.

\vskip 3mm

\enlargethispage*{2\baselineskip}
Under the assumption (\ref{ac}), the equation for the evolution of $\Phi(V;t)$ takes
the form:
\be\label{eqPhi}
\partial_t\Phi(V;t)=-{\partial \over \partial V}[\Phi(V;t)F(V;t)]
\ee
where
\be
F(V;t)=\gamma\left[\int_{V}^{\infty}(v-V)^2 a^{-}(v;V;t)dv
-\int_{-\infty}^{V}(v-V)^2 a^{+}(v;V;t)dv\right]
\ee
is continuous in $V$ for all fixed $t$.
\vskip 3mm
\noindent
\underline{\it Property 1:}
\vskip 1mm
Let $V(t)=\tau_tV_0$ be the solution of 
\be\label{SD}
\left\{
\begin{array}{l}
\displaystyle{\frac{d}{ dt }V}=F(V;t)\\
\\
V(0)= V_0
\end{array}
\right.
\ee
then the solution of (\ref{eqPhi}) is
\be\label {solPhi}
\Phi(V;t)=\Phi(\tau_{-t}V;t=0)\;{d\over dV}(\tau_{-t}V)
\ee

\vskip 3mm
\noindent
\underline{\it Proof:}
\vskip 1mm
For $t=0$, (\ref{solPhi}) is an identity.
For $t>0$,  (\ref{solPhi}) means that for any test-function $g(V)$:
\be\label{Phig}
<\Phi_t;g>=\int \Phi(V;t)g(V)dV=\int \Phi(V;0)g(\tau_{t}V)dV
\ee
From Eqs. (\ref{SD} - \ref{Phig})
follows that:
\be
\begin{array}{rl}
\displaystyle{{d\over dt}}<\Phi_t;g>&=\int \partial_t\Phi(V;t)g(V)\;dV\\
& \\
&=\int\Phi(V;0)\left({d\over dV^{\prime}}g\right)(
V^{\prime}=\tau_tV)F(\tau_tV;t)\;dV\\ & \\
&=\int \Phi(V^{\prime};t)\left({d\over dV^{\prime}}g\right)
(V^{\prime})F(V^{\prime};t)\;dV^{\prime}  \\& \\
&=-\int {\partial\over \partial V^{\prime}}
\left[\Phi(V^{\prime};t)F(V^{\prime};t)\right]\;g(V^{\prime})\;dV^{\prime}
\end{array}
\ee
i.e.
\be
{d\over dt}\Phi(V,t)=-{\partial \over \partial V}[\Phi(V;t)F(V;t)]
\ee
Therefore the distribution $\Phi(V;t)$ defined by (\ref{solPhi})
is solution of  Eq. (\ref{eqPhi}).
Let us remark that the continuity of $F(V;t)$ in $V$ is
essential in order that $\Phi(V;t)F(V;t)$ is a well-defined distribution.\\
From Property 1, we obtain immediately the following corollary.
\vskip 3mm

\noindent
\underline{\it Corollary 1:}
\vskip 2mm
If 
\be \Phi(V;t=0)=\delta(V-V_0)
\ee
 then 
\be \Phi(V;t)=\delta(V-V(t))
\ee
where $V(t)$ is the solution of (\ref{SD}). Moreover, from
Eq. (\ref{ac}) we have 
\be \rho_{surf}^{\pm}(v;V;t)=\rho_{surf}^{\pm}(v;t)\delta(V-V(t))
\ee
where \be
\rho_{surf}^{\pm}(v;t)=a^{\pm}(v; V(t);t)\ee
and
\be
\int \rho_{surf}^{\pm}(v;t)dv=\rho_{surf}^{\pm}(t)
\ee
It follows from the corollary  that for the initial condition:
\be
\Psi(X,V;t=0)=\delta(X-X_0)\delta(V-V_0)
\ee
the evolution of the piston is deterministic in the sense that
\be\label{evolpsi} 
\Psi(X,V;t)=\delta(X-X(t))\delta(V-V(t))
\ee
where $X(t)$ is obtained by integration of
$V(t)$ with $X(t=0)=X_0$.

\vskip 3mm
As for  (\ref{ac})  it is natural to assume that the distributions
$\rho_{1,P}$ and $\rho_{P,1}$ are absolutely continuous with
 respect to $\Psi(X,V,t)$, i.e.
\be
\begin{array}{rl}\label{rho1P}
\rho_{1,P}(x,v;X,V;t)&=b^-(x,v;X,V;t)\Psi(X,V;t)
\\
\rho_{P,1}(X,V;x,v;t)&=b^+(X,V;x,v;t)\Psi(X,V;t)
\end{array}
\ee
\\ 
where $b^{\pm}$ are continuous functions in $(x,X,V)$.
We obtain from (\ref{rho1P}), (\ref{16}) and (\ref{17})\\
\be\label{49}
\begin{array}{rl}
\rho_{1,P}(x,v;X,V;t)&=\rho^-(x,v;t)\Psi(X,V;t)\\
\\
\rho_{P,1}(X,V;x,v;t)&=\rho^+(x,v;t)\Psi(X,V;t)
\end{array}
\ee
where 
\be \rho^{\pm}(x,v;t)=b^{\pm}(x,v;X(t),V(t);t)\ee
 Moreover, from (\ref{29}), (\ref{30}), (\ref{49}), we obtain for
the joint distribution for the piston and particle velocity
\be\label{50}
\begin{array}{rl}
\rho_{surf}^{\pm}(v;V;t)=\rho_{surf}^{\pm}(v;t)\delta(V-V(t))
\end{array}
\ee
where 
\be 
\rho_{surf}^{\pm}(v;t)=\rho^{\pm}(X(t),v;t)=\rho_{surf}^{\pm}(t)
f^{\pm}(v;t)
\ee
and $f^{\pm}(v;t)$
is the probability distribution for  particles at the left and right surface of the piston.

\vskip 3mm
In conclusion, in the thermodynamic limit for the piston ($M = \infty$,
taken as described above), the motion of the piston is purely deterministic and the correlation functions for one particle and the piston have the factorization property (\ref{49}).

\noindent Introducing
\be\label{53}
F_2[V;\rho_{surf}^{\pm}(.)]=\int_{V}^{\infty}(v-V)^2 
\rho_{surf}^{-}(v;t)dv
-\int_{-\infty}^{V}(v-V)^2 \rho_{surf}^{+}(v;t)dv
\ee
we have thus the following equations for the evolution:
\be
{dV\over dt}=\gamma\; F_2[V;\rho_{surf}^{\pm}(.)]\label{r69}
\ee
\vskip 3mm $
(\partial_t+v\partial_x)\rho^-(x,v;t)=\delta (x+L)v\rho^-(-L,v;t)$
\be\label{54}
+\delta(x-X(t))[V(t)-v][\theta(V(t)-v)
\rho^-(X(t), 2V(t)-v;t)+\theta(v-V(t))\rho^-(X(t), v;t)]
\ee
\vskip 3mm $
(\partial_t+w\partial_y)\rho^+(y,w;t)=-\delta (y-L)w\rho^+(L,w;t)$
\be\label{55}
-\delta(y-X(t))[V(t)-w][\theta(w-V(t))
\rho^+(X(t), 2V(t)-w;t)+\theta(V(t)-w)\rho^+(X(t), w;t)]
\ee
It follows from (\ref{54}) (\ref{55})   that:
\begin{eqnarray}
\rho^-(x,v;t)&=&\rho^-(x_0,v_0;t=0)\\
\rho^+(y,w;t)&=&\rho^+(y_0,w_0;t=0)
\end{eqnarray}
where $(x_0,v_0)$, resp. $(y_0,w_0)$, is the initial conditions
necessary to reach the point $(x,v)$ (resp. $(y,w)$) at time $t$, 
under the free evolution of the particle with elastic collisions at the boundary $x=-L$
and $x=X(t)$ (resp $y=X(t)$ and $y=L$)

Finally, from (\ref{50}) and (\ref{49}), the assumptions (\ref{ac})
and (\ref{rho1P})  will be satisfied, so that the 
whole reasoning is consistent.

\clearpage

Let us summarize our results in the following statement:

\noindent
\underline{\it Property 2:} 
\vskip 1mm
In the thermodynamic limit for the piston, and
for the initial condition
$$\Psi(X,V;t=0)=\delta(X-X_0)\delta(V-V_0)$$
\noindent
the evolution of the piston and the gases is described by the autonomous system of equations (\ref{r69} - \ref{55}). The evolution of the piston is deterministic, $i.e.$
\be
\Psi(X,V;t)=\delta(X-X(t))\,\delta(V-V(t))\label{74b}
\ee
and the joint probability distributions for one particle and the piston have the factorization property
\be
\rho^{\pm}(x,v;X,V;t)=\rho^{\pm}(x,v;t)\,\Psi(X,V;t).\label{75b}
\ee
\vskip 5mm
\noindent
\underline{\it Thermodynamic quantities on the surface of the piston}
\vskip 1mm
We introduce $F_2^{\pm}(V)$, functionals of $\rho^{\pm}_{surf}(v;t) $ defined by
\begin{eqnarray}
F_2^-(V)&=&\int_{V}^{\infty}(v-V)^2 
\rho_{surf}^{-}(v;t)dv\\
&&\\
F_2^+(V)&=&
\int_{-\infty}^{V}(v-V)^2 \rho_{surf}^{+}(v;t)dv
\end{eqnarray}
In the thermodynamic limit, $\gamma$ simply writes
 $\gamma=2mA/M$ and the 
equation of motion (\ref{r69}) is of the form
\be\label{61}
{d\over dt}V={A\over M}\;[2mF_2^-(V)-2mF_2^+(V)]
\ee
where $2mF_2^-(V)$ is the force per unit area exerted by the left particles
on the piston and similarly $-2mF_2^+(V)$ is the force per unit area
exerted by the right particles.
Since $F_2^-(V)$ and $-F_2^+(V)$ are monotonically decreasing functions of $V$
(for any distribution $\rho_{surf}^{\pm}(v)$),
we are led to define the {\it pressure} $\widehat{p}$,
the {\it friction coefficients} $\lambda(V)$ as well
as the {\it density} $\widehat{\rho}$ and the 
{\it temperature}  $\widehat{T}$ associated with the particles which are going to hit  the piston
by
\begin{eqnarray}
\widehat{\rho}\;^-&\doteq& 2\int_0^{\infty}\rho_{surf}^{-}(v;t)dv\label{62}\\
&&\nonumber\\
\widehat{\rho}\;^+&\doteq& 2\int_{-\infty}^{0}\rho_{surf}^{+}(v;t)dv\label{62bis}\\
&&\nonumber\\
\widehat{p}\;^{\pm}&\doteq &2mF_2^{\pm}(V=0)\doteq 
\widehat{\rho}\;^{\pm}k_B\widehat{T}\;^{\pm}\label{63}\\
&&\nonumber\\
2mF_2^{\pm}(V)&\doteq &\widehat{p}\;^{\pm}\pm\;
{M\over A}\;\lambda^{\pm}(V)V\label{64}
\end{eqnarray}
where $\lambda^{\pm}(V)$ is positive for all $V$.
The equation (\ref{61}) for the evolution of the piston is thus
\be\label{65}
{d\over dt}V={A\over M}\;(\widehat{p}\;^--\widehat{p}\;^+)
-[\lambda^{-}(V)+\lambda^{+}(V)]V
\ee
and involves the difference of pressure and a ``friction force''.

We notice that for distributions $\rho_{surf}^{\pm}(v;t)$
which are symmetric in $v$, (\ref{62}), (\ref{62bis}) and (\ref{63})
are the standard definitions of density, pressure and temperature.
 However this symmetry property is not satisfied in our case.

Moreover in the thermodynamic limit  for the piston we also have from
(\ref{54}), (\ref{55})
\begin{eqnarray}
{d\over dt}\left({<E^->\over A}\right)&=& -2mF_2^-(V)V\label{66}\\
&&\nonumber\\
{d\over dt}\left({<E^+>\over A}\right)&=& 2mF_2^+(V)V\label{66bis}
\end{eqnarray}
where $<E^{\pm}>/A$ is the (kinetic)  energy per unit area of the particles
in the left and in the right compartment. From the first principle of
 thermodynamics we thus have 
the following result.

\vskip 3mm
\noindent
\underline{\it Property 3:}
\vskip 1mm
In the thermodynamic limit the evolution of the system is ``adiabatic'',
i.e. there is only work and no heat involved.
\vskip 3mm
\noindent
\underline{\it Remarks}

\begin{enumerate}
\item It should be stressed that the deterministic evolution (\ref{74b}),
 the factorization property (\ref{75b}), and the adiabatic
 property (\ref{66}-\ref{66bis}) are only valid in the thermodynamic limit.
 For systems with finite mass $M$, these properties 
will only be approximately verified for a time interval of the order of $M$ \cite{suite}. 
\item The system of equations (\ref{r69}-\ref{55}) which we have obtained is  identical to the equations derived in  \cite{LPS} using a different point of view. Indeed, as mentionned in the introduction, the authors have considered in  \cite{LPS} another thermodynamic limit in wich $L$ is not fixed but tends to infinity, together with $M_L \sim L^2$, $N^{\pm}_L\sim L^3$. Position is then scaled with $y = L^{-1}\, x$ and time is scaled with $\tau = L^{-1}\, t$. However they also introduce an assumption on the distribution functions which is equivalent to a scaling on the number of particles with $\widetilde{N} = L^{-1}\,N^{\pm}_L \,\sim \,L^2$. Therefore it is not surprising that their equations in terms of the scaled variables is identical to the equations we have obtained above.
\end{enumerate}
\section*{ 4. Evolution of the piston in the thermodynamic limit ($M=\infty$)
and for the infinite cylinder ($L=\infty$)}

In this section, we consider the limiting situation where the length of 
the cylinder is infinite and the thermodynamic limit for the piston
has been taken (Sec. 3)

Assuming that the initial distributions $f^{\pm}$(v) are zero
if $|v|\not\in [v_{min}, v_{max}]$, where $v_{min}$
and $v_{max}$ depend on the initial temperatures and pressures,
there will be no recollision of the particles on the piston. In
this case $\rho_{surf}^{-}(v;t)$, resp $\rho_{surf}^{+}(w;t)$,
will be independent of $t$ for all
$v>\inf_t V(t)$, resp. $w<\sup_t V(t)$, and thus $F_2(V)$
is independent of $t$. If the initial distributions are
Maxwellian with $|p^--p^+|$ sufficiently small, the velocity of 
piston will be small and we also expect  that $F_2(V)$ is approximately independent
of $t$. In those cases ${\rho}^{\pm}(v;t)=\rho^{\pm}(v)$ for $v$  {\footnotesize\raisebox{-1mm}{$\stackrel{{\displaystyle{<}}}{>} $}} $\,\, 0$, 
 $\widehat{T}^{\pm}(t)=T^{\pm}$ and $\widehat{p}\;^{\pm}(t)= p^{\pm}$.

The evolution of the piston is thus simply given by the ordinary differential
equation
\vspace{3mm}
\be\label{67}\vspace{1mm}
\left\{
\begin{array}{rl}
{d\over dt}V&=\left({A\over M}\right)\;2mF_2(V) \\
&\\
V(0)&=0
\end{array}
\right.
\ee\vspace{1mm}
where $F_2(V)$ is a strictly decreasing function of$V$:
\be\label{68}
2mF_2(V)=(p^--p^+)-\frac{M}A\lambda(V)V\hspace{10mm}
\lambda(V)=\lambda^+(V)+\lambda^-(V)>0
\ee
If the initial conditions are such that $p^-=p^+$
then from (\ref{67}), (\ref{68}), we have  $V(t)=0$ for all $t$, i.e. the
piston remains at rest and the gases on both  sides remain in equilibrium
with their respective temperatures $T^-$ and $T^+$: the piston is therefore
an adiabatic wall in the thermodynamical sense.

If the initial conditions are such that $p^-\neq p ^+$
(i.e. $\rho^-k_BT^-\neq\rho^+k_BT^+$), then from  (\ref{67}), (\ref{68}), 
the piston will evolve monotonically (i.e. either $X(t)$ increases, or decreases, 
for all $t$) to  the stationary state with constant velocity $\bar{V}$
solution of the equation $F_2(\bar{V})=0$, i.e.
\be\label{69}
\int_{\bar{V}}^{\infty}\rho^-(v)(v-\bar{V})^2dv
-\int_{-\infty}^{\bar{V}}\rho^+(v)(v-\bar{V})^2dv=0
\ee
Moreover, the approach to the stationary state is exponentially fast with 
time constant $\tau_0$
\be\label{70}
\tau_0={1\over \lambda}\hspace{10mm}
{\rm where}\hspace{10mm}\lambda=\lambda(V=0)
\ee
Assuming that for $k=1$ and 3, the expressions
\be
\int_0^{\infty}v^k\rho^-(v)dv\hspace{10mm}
{\rm and}\hspace{10mm}\int_{-\infty}^0v^k\rho^+(v)dv
\ee
are approximately the same functions of $T^-$ and
$T^+$ as for the Maxwellian distributions, we have
\be\label{71}
\tau_0^{-1}=\lambda=\frac{A}M\sqrt{8k_Bm\over\pi}
\;(\rho^-\sqrt{T^-}+\rho^+\sqrt{T^+})
\ee
and $\bar{V}$ is given by the solution (closest to 0) of
$$\hspace*{-40mm}k_B\;(\rho^-T^--\rho^+T^+) -\bar{V}\;\sqrt{8k_Bm\over\pi}\;
(\rho^-\sqrt{T^-}+\rho^+\sqrt{T^+})$$
\be\label{Vbar}\hspace*{20mm}
+\bar{V}^2m(\rho^--\rho^+)-\bar{V}^3\;{m\over 6}\;
\sqrt{8m\over\pi k_B}\;\left({\rho^-\over\sqrt{T^-}}+{\rho^+\over\sqrt{T^+}}
\right) +{\cal O}(\bar{V}^4)=0
\ee
In conclusion, the stationary velocity $\bar{V}$ is a function of
$\rho^+/\rho^-$, $T^+$, $T^-$ and $m$, but does not depend
on the values $M/A$, $N^{\pm}/M$. Furthermore the time necessary to reach the stationary
velocity is characterized by $\tau_0$, which from (\ref{70}), (\ref{71})
tends to zero when $N^{\pm}/M$ is very large
(with fixed $T^-$, $T^+$).
In Sec. 6, we will check the above conclusions
(in fact the assumptions under which they
have been derived) by means of numerical simulations.
 
To conclude this section, we remark that in the thermodynamic
limit the stationary state is stable with respect
 to small perturbations of $\bar{V}$. We should also remark that 
for small $\bar{V}$
and Maxwellian distributions for the velocities,
 the equation (\ref{69}) is similar to the equation obtained from fluid dynamics under the assumption that the evolution is adiabatic \cite{ll}.

\section*{ 5. Evolution of the piston in the thermodynamic limit ($M=\infty$)
and for a finite cylinder ($L<\infty$)}

If the cylinder has finite length, we have to consider the full set of
equations (\ref{53}-\ref{55}). In this section, we shall
give only a qualitative discussion and then in Sec. 6,
we shall compare our conclusions with the numerical
simulations. To simplify the notation, as in the simulation we now place the left 
boundary at $x=0$ and the right boundary at $x=L$, i.e. the cylinder
has now a finite length $L$.

Introducing the average temperatures $T^{\pm}$ of the gases
 in the left and right compartments by
\be
T^{\pm}={2<E^{\pm}>\over k_BN^{\pm}}
\ee
the conservation of energy implies in the thermodynamic limit for the
piston (recall that $<V^2>=<V>^2$ in this limit)
\be\label{74}
\left({N^-\over N}\right)T^-_t+\left({N^+\over N}\right)T^+_t
+{M\over Nk_B}V^2_t=cte\doteq T_0
\ee
where $N=N^++N^-$.
From Eqs. (\ref{61},\,\,\ref{66},\,\,\ref{66bis}), we then have
\be\label{75}
{d\over dt} V=\left({A\over M}\right) \;2m\;[F_2^-(V)-F_2^+(V)]
\ee
\be
k_B{d\over dt} T^-=-4m\left({A\over N^-}\right) \;F_2^-(V)V
\ee
\be
k_B{d\over dt} T^+=4m\left({A\over N^+}\right) \;F_2^+(V)V
\ee
where 
\be\label{78}
2mF_2^{\pm}(V)=\widehat{\rho}^{\pm}k_B\widehat{T}^{\pm}
\pm\left({M\over A}\right) \;\lambda^{\pm}(V)V
\ee
We notice that  (\ref{74}) is satisfied fo all $t$.
Let us first remark that in (\ref{75}), appears 
$m(A/M)\widehat{\rho}_{surf}^{\pm}$ which at $t=0$ is
$R^-/X_0$ and $R^+/(L-X_0)$ with
\be
R^{\pm}=m\;{N^{\pm}\over M}
\ee
Therefore, as discussed in Sec. 4, if $R^{\pm}$ is sufficiently
large then after a time $\tau_0={\cal O}((R^{\pm})^{-1})\ll 1$
the piston will reach the velocity 
$\bar{V}=\bar{V}(\rho^+/\rho^-, T^{\pm})$ given by Eq. (\ref{69}),
i.e. $X(t)=X_0+\bar{V}t$.
This evolution will continue over a time interval of the order
\be \tau_1=\min\left\{
{ 2 X_0\over\sqrt{\frac{ 3_BT^-}m}-\bar{V}}, {2(L-X_0)\over\sqrt{\frac{3_BT^+ }{m}+\bar{V}}}
\right\}
\ee
which is the time for the sound wave to propagate from the piston 
to the closest boundary of the cylinder.\\
For $t>\tau_1$, the density distributions $\rho_{surf}^{\pm}(v;t)$, $v $ {\footnotesize\raisebox{-1mm}{$\stackrel{{\displaystyle{<}}}{>} $}} $\,\, 0$
depend explicitly on time. To discuss the evolution we now introduce the following

\vskip 3mm
\noindent
\underline{\it Average assumption:}
\vskip 1mm
\noindent\noindent The thermodynamic quantities associated with the surface of the piston (\ref{62}-\ref{64}), defined by $\rho_{surf}^{\pm}(v;t)$ with $v \,\,  ${\footnotesize\raisebox{-1mm}{$\stackrel{{\displaystyle{>}}}{<} $}} $\,\, 0$, 
are approximately equal to the average of the corresponding quantities in the left/right compartement, $i.e$
\be\label{hom1}
\widehat{\rho}^-(t)\simeq  {N^-\over AX(t)}
\hspace{20mm}
\widehat{\rho}^+(t)\simeq  {N^+\over A(L-X(t))}
\ee
\be\label{hom2}
\widehat{T}^{\pm}(t)\simeq T^{\pm}(t)
\ee
Let us remark that this average assumption is usually introduced in the experimental measurements of  the ratio of the specific heats of gases, $\gamma= c_p/c_v$ [4,18].\\
With this assumption, Eqs. (\ref{75}-\ref{78}) yield
\be\label{83}
{d\over dt}V={N^-\over M} k_B{T^-\over X} -{N^+\over M}k_B{T^+\over (L-X)}
-\lambda(V)V
\ee
\be\label{84}
{d\over dt}T^-=-2T^-{V\over X}+2V^2\left({M\over N^-}\right)\;
{\lambda^-(V)\over k_B}
\ee
\be\label{85}
{d\over dt}T^+=2T^+{V\over L- X}+2V^2\left({M\over N^+}\right)\;
{\lambda^+(V)\over k_B}
\ee
(once again one can see that (\ref{74}) is satisfied for all $t$).
Therefore introducing
\be
\sigma^-=\sqrt{T^-}X
\hspace{10mm}{\rm and}\hspace{10mm}
\sigma^+=\sqrt{T^+}(L-X)
\ee
eqs. (\ref{83}-\ref{85}) implies
\be\label{87}
{d\over dt}\sigma^-=\left({M\over A}\right)\;
{\lambda^-(V)\over \rho^-k_B\sqrt{T^-}}\;V^2
\ee
\be\label{88}
{d\over dt}\sigma^+=\left({M\over A}\right)\;
{\lambda^+(V)\over \rho^+k_B\sqrt{T^+}}\;V^2
\ee
\vskip 5mm
Let us note that the entropy of the ``one-dimensional'' perfect
gases is
\be\label{108}
{S^-\over A}=\left({N^-\over A}\right)\;
k_B \;\ln[\sqrt{T^-}X]+f(N^-/A)
\ee
\be\label{109}
{S^+\over A}=\left({N^+\over A}\right)\;
k_B \;\ln[\sqrt{T^+}(L-X)]+f(N^+/A)
\ee
Therefore from (\ref{87}), (\ref{88}),
\be
{d\over dt}\left({S^{\pm}\over A}\right)
=\left({M\over A}\right)\;
{\lambda^{\pm}(V)\over T^{\pm}}\;V^2
\ee
and  we have thus obtained the following

\vskip 3mm
\noindent
\underline{\it Property 4:}
\vskip 3mm
Under the average assumption (\ref{hom1}-\ref{hom2}),
the evolution of the piston is adiabatic, i.e. no heat transfer is involved 
and the entropy of both gases defined by (\ref{108}),(\ref{109}) are strictly increasing in time.

Moreover, assuming that 
\be
\int_0^{\infty}v\rho_{surf}^-(v;t)dv\hspace{10mm}
{\rm and}\hspace{10mm}\int_{-\infty}^0v^k\rho_{surf}^+(v;t)dv
\ee
are approximately the same functions of $T^-$ and
$T^+$ as for the Maxwellian distributions, we have (\ref{71})
\be\label{89}
\lambda^{\pm}(V=0)=\left({A\over M}\right)\;\sqrt{8k_Bm\over\pi}
\;\rho^{\pm}\sqrt{T^{\pm}}
\ee
i.e.
\be
\lambda^{-}(V=0)=\left({N^-\over M}\right)\;\sqrt{8k_Bm\over\pi}
\;{\sigma^-\over X^2}
\ee
\be
\lambda^{+}(V=0)=\left({N^+\over M}\right)\;\sqrt{8k_Bm\over\pi}
\;{\sigma^+\over (L-X)^2}
\ee
In conclusion, from (\ref{87}), (\ref{88}) and (\ref{89})
we have
\begin{eqnarray}
\label{117}
&&{d\over dt}(\sigma^--\sigma^+)={\cal O}(V^3)\\
\label{118}
&&{d\over dt}V=k_B\left[
\left({N^-\over M}\right)\;{(\sigma^-)^2\over X^3}
- \left({N^+\over M}\right)\;{(\sigma^+)^2\over (L-X)^3}
\right]-\lambda(V)V
\end{eqnarray}
and thus if the evolution is such that the velocity is small for all $t$, then
\be
\sigma^--\sigma^+\cong cte
\ee
\vskip 3mm
\noindent
\underline{\it Equilibrium point \ ($M=\infty$):}
\vskip 1mm
From Eqs. (\ref{74},\ref{83}-\ref{85}) we find that the equilibrium point
is given by
\begin{eqnarray}\label{mecheq1}
\left({N^-\over N}\right)T_f^-&=&T_0\;{X_f\over L}\\
\label{mecheq2}
\left({N^+\over N}\right)T_f^+&=&T_0\left(1-\frac{X_{f}}{L}\right)
\end{eqnarray}
i.e. in particular $p_f^-=p_f^+$. Furthermore from 
(\ref{117}) taking $\sigma^--\sigma^+\approx cte$
yields
\be\label{mecheq3}
\sqrt{T^-}X-\sqrt{T^+}(L-X)=C\doteq
\sqrt{T^-(0)}X_0-\sqrt{T^+(0)}(L-X_0)
\ee
i.e. with (\ref{mecheq1}), (\ref{mecheq2})
\be\label{mecheq4}
\sqrt{\left({N\over N^-}\right)X_f^3}
-\sqrt{\left({N\over N^+}\right)(L-X_f)^3}=\sqrt{L\over T_0}\;C
\ee
Solving (\ref{mecheq4}), (\ref{mecheq1}), 
(\ref{mecheq2}) with the constants $C$, $T_0$ given by the initial
conditions (\ref{74}), gives
the equilibrium state $(X_f, T_f^-, T_f^+)$ which is a state
of mechanical equilibrium $p_f^-=p_f^+$
but not thermal equilibrium $T_f^-\neq T_f^+$.
It is important to realize that thermostatics yields only the condition of equality of the pressures, $i.e.$ mechanical equilibrium. In our dynamical approach, we have obtained a new equation (\ref{117}), consistent with the second principle of thermodynamics ({\it property $4$ }) which then enables  us to determine the equilibrium point.

\vskip 3mm
\noindent
\underline{\it Linearization around the equilibrium point:}
\vskip 1mm
Let $(X_f, T_f^{\pm})$ be the equilibrium point given by
(\ref{mecheq1}-\ref{mecheq4}) and $x=X-X_f$.
Linearizing the equations (\ref{83}), (\ref{87}), (\ref{88}),
together with (\ref{89}) for the friction coefficient, yields
\be
\left\{
\begin{array}{l}
\sigma^-=cte=\sqrt{T_f^-}X_f\\
\sigma^+=cte=\sqrt{T_f^+}(L-X_f)\\
\ddot{x}=-\omega_0^2x-\lambda\dot{x}
\end{array}
\right.
\ee
with
\be\label{99}
\omega_0^2=3\;\left({N\over M}\right)\;
{k_BT_0\over X_f(L-X_f)}
\hspace{20mm}N=N^++N^-
\ee
\be
T_0=\left({N^-\over N}\right)\;T^-(0)
+\left({N^+\over N}\right)\;T^+(0)
\ee
\be
\lambda=\sqrt{8m\over\pi}
\;\sqrt{{N\over M}\;{k_BT_0\over L}}
\left[\sqrt{{N^-\over M}\;{1\over X_f}}+
\sqrt{{N^+\over M}\;{1\over (L-X_f)}}
\right]
\ee
In conclusion the equilibrium point (for the thermodynamic limit)
is stable and the approach to equilibrium is a damped, harmonic, oscillation

\be
X(t)= C_+ e^{\displaystyle{-\varrho^+ t}} + C_- e^{\displaystyle{-\varrho^- t}}, \qquad\quad\varrho^{\pm}={\lambda\over 2}\mp
\sqrt{{\lambda^2\over 4}-\omega_0^2}
\ee
Moreover, in the linear approximation the evolution of the gases
are at constant entropy.
 \\
\vskip 0.1mm
\noindent
\underline{\it Remarks}
\begin{enumerate}
\item Although our derivation of the evolution equation was strictly 
Hamiltonian (Liouville equation), and the concept 
of entropy was never introduced, it is interesting  to note
 that the frequency $\omega_0$, (\ref{99}), which we have obtained
 with our assumption coincides with the frequency  obtained in 
thermodynamics assuming ``adiabatic oscillations" [4,18].
\item

For the case $N^+=N^-=\bar{N}$ which will be considered in the simulations
(Sec. 6)
\be
R={\bar{N}m\over M}={M^{\pm}\over M}
\ee
\be
T_0={1\over 2}\;(T^-(0)+T^+(0))
\ee
\be\label{105}
\lambda=\sqrt{8R\over 3\pi}\;
\left[\sqrt{X_f\over L}+\sqrt{1-{X_f\over L}}\right]\;\omega_0
\ee
This implies that the motion is weakly damped if
\be
R\leq R_{max}={3\pi\over 2}\left[
\sqrt{X_f\over L}+\sqrt{1-{X_f\over L}}
\right]^{-2}
\ee
with ``period''
\be
\tau={2\pi\over \omega_0}\;{1\over \sqrt{1-R/R_{max}}}
\ee

\end{enumerate}

We have thus obtained the conclusion that  if $R<2.3$ the motion
 is weakly damped, and if $R>4.7$ the motion is strongly damped.
 Again we should mention that the main result of the 
recent experimental measurements of $\gamma$ is that one 
should distinguish between two regimes, corresponding 
to weak and strong damping with very different properties [18].
 In particular, experimental results show that the frequency of oscillations for weak damping is very close to the values obtained assuming adiabatic oscillations, eq.(122).
 
\section*{6. Numerical simulations}

To illustrate the results of Sec. 4 and to check
the conclusions of Sec. 5 which were based on the average assumptions 
(\ref{hom1}), (\ref{hom2}), we have performed a large number
 of numerical simulations.
In all simulations, we have considered a one-dimensional
system of fixed length $L$, with the left boundary at $x=0$, and we have taken:
\be
k_B=1, \hspace{5mm}m=1,\hspace{5mm} L= 60 \cdot 10^4, \hspace{5mm}X_0=10\;.\;10^4,\hspace{5mm} V_0=0
 \ee
\be N^-=N^+=\bar{N},\hspace{5mm}R={\bar{N}\over M},\hspace{5mm}{\rm i.e.}\;\;\;
\rho^-(0)=5\rho^+(0)\ee
A very large number of simulations has been conducted with the number $\bar{N}$ of particles in the left/right compartments ranging from $10^3$ to $500 \cdot 10^3$, the mass $M$ of the piston from  $1$ to $ 10^5$, and the parameter  $R$ from $0.1$ to $ 10^3$.

At the initial time $t=0$, the left and the right particles
are taken with Poisson distribution for the position
and Maxwellian distribution for the velocities, characterized by:
\begin{eqnarray}
\label{136}
T^-=1, \hspace{5mm}T^+=10, \hspace{5mm}{\rm i.e.}\;\;\;T_0=5.5
\end{eqnarray}
and thus
\begin{eqnarray} 
\label{137}
p^+=2p^-
\end{eqnarray}
Since under the scaling $L'=\alpha L,\;\; X_{0}'=\alpha X_{0}$, time
is scaled with $t'=\alpha t$, we have kept $L$ fixed. Similarly under the scaling 
$T'^{\pm}=\alpha T^{\pm}$, time is scaled with $t'=
\frac{1}{\sqrt{\alpha}}t$, and thus  we have also kept $T^{-}$ and $T^{+}$ fixed.

From the discussion of Section 4, it is expected that for $R$ 
sufficiently large, the actual velocity of the piston before the first
 recollisions of the particles will coincide with 
the stationary velocity $\bar{V}$, given by eq.(\ref{69}), 
for the infinite cylinder and the initial conditions
 (\ref{136}-\ref{137}). 
 For the parameters used in the simulations eq.($86$) yields
\begin{eqnarray}
\label{112}
\bar{V}=-0.3433
\end{eqnarray}
The velocity $c$ of the sound wave in the one-dimensional 
perfect gas is given by $c=\sqrt{3k_BT/m}$.
Therefore for our simulations, the time needed for the sound wave
to make the first collision on the piston is:
\be\label{113}
t^-={2X_0\over c^--\bar{V}}=0.96\;.\;10^5\hspace{15mm}
t^+={2(L-X_0)\over c^++\bar{V}}=1.95\;.\;10^5
\ee
The mechanical equilibrium defined by eqs.
(\ref{mecheq1}-\ref{mecheq4}) is:
\begin{eqnarray}
\label{140}
X_f&=&8.42\;.\;10^4\label{114}\\
T_f^-&=&1.54\hspace{15mm}T_f^+=9.46\label{115}\\
p^-_f&=&p^+_f=1.83(4)\;\bar{N}\;.\;10^{-5}\label{116}
\end{eqnarray}
For the approach to equilibrium, 
we have from eqs. (\ref{99}), (\ref{105}):
\begin{eqnarray}
\label{143}
\omega_0=\sqrt{R}\;.\;0.2756\;.\;10^{-4},
\hspace{10mm}\lambda\cong R\;.\;0.331\;.\;10^{-4}
\end{eqnarray}
and the ``period'' of the damped oscillations is:
\be
\tau=2.28\;.\;10^5\;[R(1-0.36 R)]^{-1/2}
\ee
which yields:
\be
R_{max}\cong 3\ee Let us remark that for thermal equilibrium
of a heat-conducting piston, one would have:
\be
X_{th}=30\;. \;10^4, \quad T_{th}^-=T_{th}^+=5.5,\quad
p^-=p^+=1.833\;\bar{N}\;.\;10^{-5}\ee
In figure $1$, we have taken $R=4$ and investigated 
the thermodynamic limit by considering  increasing
values of $M$ from $10\;000$ to $100\;000$.
From figure 1, we conclude that for $t<30\;.\;10^5$, the thermodynamic
limit for the piston is reached when $M>50\;000$.
Other simulations with fixed $R$ ranging from 0.1 to 300 confirm this result:
for $t<30.10^5$, the evolution corresponds to the thermodynamic limit if
$M>50\;000$.
\vskip 3mm
In figures $2$ and $3$, we have  considered the time evolution for the piston 
for different values of $R$, and values of $M$ such that the evolution 
corresponds to the thermodynamic limit.
The simulations $(Fig. 2)$ show that the evolution is very weakly
damped for $R<1$ and strongly damped for $R>4$ in agreement with the 
conclusion of Sec. $5$. Moreover the piston evolves toward the equilibrium position $X_f \cong 8.35 \cdot 10^4$, which is in excellent agreement with the predicted value $X_f =8.42 \cdot 10^4$ obtained from the result  of section $5$, $i.e.$ eq.(140).

The weakly damped oscillations have a period
which is in very good agreement  with the expected values
 of Sec.$ 5$ (see Table$ 1$),
but the damping coefficient computed over more than $1000$ oscillations gives
a value several order of magnitude smaller than  $\lambda$ 
given by eq.(\ref{143}). For example for $R=0.2,\;\;M=100\,000$, we observe a damping equal to $3.10^{-5}\lambda$. 
The oscillations occur around the equilibrium position $X_f$, and their period remains constant in time (at least up to $t=7000 \cdot 10^{5}$ which corresponds to $1500$ oscillations and is the time the simulation has been running). As predicted, the frequency of the oscillations increases with $R$, but is independent of $M$ for sufficiently large $M$. The fluctuations on the amplitude of these oscillations is of order $1\%$.

On the other hand for $R>10$ and $M>6\;000$, i.e. for strongly damped evolution, it is seen on figure  $3$ that the evolution is independent of $R$ and $M$
 if $t<20.\;10^5$.
We also  notice (Fig. 3) that for $R>10$, the piston acquires almost immediately a velocity:
\be
V_1=-0.34
\ee
which is in perfect agreement with $\bar{V}$ computed above with
 eq. (\ref{112}).
Therefore for strong damping the friction coefficient coincide with 
the computed value during the initial evolution, $i.e.$ as long as the influence of the boundaries ($x=0,\;x=L$) did not appear.
The piston then arrives at its first minimum position at the time
$t_1=0.9\;.\;10^5$, which corresponds to the time $t^-$ for the sound 
wave to return on the piston, eq. (\ref{113}), and 
reaches the position of mechanical equilibrium after two oscillations, $i.e.$ 
at a time $t$ which is about $4t_1$ (see also Fig. 4).
It is seen on figure $1$ and $2$ that for $R$ between $4$ and $10$ new oscillations appear after a time
of the order $20\; t_1$ with amplitude
$\delta X/L\approx 4\;.\;10^{-3}$ and period
approximately equal to $2t_1$ independent of $R$ and $M$
(for $M$ large).
The amplitude of these new oscillations is modulated
 with a period about $20\;t_1$ and they tend to disappear with time.
 We are led to conjecture that these oscillations are associated with 
 sound waves propagating in the gases.
 As we have checked, these oscillations reflect the fact that
 the velocity distributions of the gases are not Maxwellian. 
They will be responsible for the approach to Maxwellian distributions. Similar results and conclusions have been obtained in [19]. For larger values of $R$ (e.g. $R=50$ or $100$ as shown in figure 3) the number of particles necessary to observe the adiabatic evolution over a time interval $40 \, t_1$ is several millions. These simulations would need a very large computer time and have not been conducted.
\vskip 3mm
In figure $4$, we show the time evolution for the temperature
$T^{\pm}$ of the gas and for the pressure $p^{\pm}$ defined by 
\be\label{111}
p^-(t)={\bar{N}k_BT^-(t)\over X(t)}\hspace{10mm}
p^+(t)={\bar{N}k_BT^+(t)\over(L- X)(t)}
\ee 
where $T^{\pm}=2E^{\pm}/\bar{N}k_B$, $E^{\pm}$ being the energies of the
two compartments.
It is seen that these evolutions follow
the evolution of the piston.
Moreover, we notice that the temperature increases (resp. decreases)
under compression (resp. expansion)
of the gas, which reflects the fact that the evolution is adiabatic.
We recall that the area $A$ does not play any role.
From  figure $4$, we obtain the numerical values for temperatures
and pressure when mechanical equilibrium is reached,
\be
T_f^-=1.52,\hspace{10mm}T_f^+=9.48,\hspace{10mm}
p^-=p^+=5.5
\ee
which are  again in perfect agreement with the expected
values (\ref{114}-\ref{116}) 
 $T_{f}^{-}=1.54, \; T_{f}^{+}=9.46,\; p_{f}=5.5$
since $\bar{N}=300\;000$.

Finally from (\ref{108}-\ref{109}), we have for the change in 
entropy between the initial and final states:
\begin{eqnarray}
\Delta S^-&=&\bar{N} k_B\;23\;.\;10^{-3}\\
\Delta S^+&=&\bar{N} k_B\;6.7\;.\;10^{-3}
\end{eqnarray}
We thus observe numerically an increase in entropy
for both gases which once more confirms the adiabatic property
of the piston.
The result observed in figure $3$  that for $R>10$
the evolution is independent of $R$ leads us to the following 

\vspace{2mm}
\noindent\noindent\underline{\it Conjecture}

\vspace{2mm}
In the thermodynamic limit, the evolution for $R\to \infty$
is given by the solution of:
\begin{eqnarray} 
(\partial_t + v \,\partial_x)\,\,\rho^- (x,v;t)&=& \delta(x) \,\,v \,\,\rho^- (0,v;t)
+ \delta(x-X(t)) \,\,[V(t)-v ]\,\,\rho^- (X(t),v;t)\nonumber \\
(\partial_t + v \,\partial_y)\,\,\rho^+ (y,v;t)&=& -\delta(y -L) \,\,v \,\,\rho^+ (L,v;t)
- \delta(y-X(t))\,\, [V(t)-v ]\,\,\rho^+ (X(t),v;t)\nonumber\\
\text{where} \quad \frac{d}{dt}X = V(t) \quad &\text{and}& \quad V(t)\quad  \text{is the solution of}\nonumber\\
\int_{V(t)}^{\infty} dv \,\rho^- (X(t),v;t)\, (v-V(t))^2& =& \int^{V(t)}_{-\infty} dv\,\rho^+ (X(t),v;t)\, (v-V(t))^2
\end{eqnarray}
This conjecture follows from (\ref{53}-\ref{55}), with the left boundary placed at the origin $(x=0)$, and the boundary condition at $x=X(t)$
\begin{eqnarray} 
\rho^{\pm} \big(X(t),v;t\big)=\rho^{\pm} \big(X(t),2 V(t)-v;t\big)\nonumber \qquad\qquad
\text{for $ v$  {\footnotesize\raisebox{-1mm}{$\stackrel{{\displaystyle{<}}}{>} $}} $\,\, V(t)$}
\end{eqnarray}
which is expected to hold in the thermodynamic limit \cite{LPS}.

\section*{ 7.Conclusions} 

As mentionned in the introduction, we have shown that in the thermodynamic
limit for the piston ($A\rightarrow\infty$ with $N^{\pm}/A$, $M/A$,
and $L$ fixed) the evolution of the piston is deterministic, i.e.
the distribution function is
$\Psi(X,V;t)=\delta(X-X_t)\delta(V-V_t)$ where $X(t)$ is the solution
of a system of autonomous equations coupled to the one-particle
distributions for the gases.
Moreover we have shown that in this limit the two-point correlation
 function for one left (right) particle with the piston 
factorizes as the product of the individual distribution functions.
Furthermore, the evolution is strictly adiabatic
(for $M=\infty$) in the sense that the changes in energy are entirely
 due to work, no heat transfer is involved and the entropy of both gases 
are strictly increasing. If the length of the cylinder is infinite
(and $M=\infty$), the piston evolves towards a stationary equilibrium
state with velocity $\bar{V}$  approximately proportional to $(p^+-p^-)$.
Therefore, in the thermodynamic limit (and $L=\infty$) 
the stationary  state is a state  of mechanical equilibrium, iff $p^+=p^-$.
\\
If the length of the cylinder is finite, we have introduced an 
assumption  to express the density of particles and the temperature
at the surfaces of the piston by their average values in the left
and right compartments. With this assumption, we were able to
analyze the evolution  and the final state.
The numerical simulations, as well as the analytical expressions,
 have shown that for finite $L$ the motion is characterized by 
damped oscillations. It is weakly damped if $N/M$ is small and 
strongly damped if $N/M$ is large. Moreover the numerical values 
obtained in the simulations confirm all the conclusions derived from 
the average assumption, except for the damping oscillations which
 appears a lot smaller than predicted. We are thus led to the conjecture 
that there are two different mechanisms responsible for friction. One mechanism correctly described by the discussion of Section 4 is associated with the motion of the piston in the absence of recollision ($L=\infty$) and is responsible for the stationary state in the infinite cylinder. Another mechanism which is responsible for the damping of oscillations appears to be associated with sound waves, i.e. inhomogeneities in the gases, traveling back and
 forth between the piston and boundaries. Finally we have observed that
 over a time interval of order $M$ the evolution is independent of $N/M$, 
as soon as $N/M$ is large enough. 
It remains however to give a proof of the result without  introducing
the 
average 
assumption.
\ \\
 In a forthcoming paper, we shall present 
numerical simulations which show the key difference between the 
infinite-mass and the finite-mass problems \cite{suite}.
For the finite-mass case, we have the following picture.
 In a first stage, characterised by a time scale of 
order  $\tau_1 = L \sqrt{M/E_0}$, the motion of the piston
 is adiabatic (no heat transfer) and corresponds to the
 motion that we have described in the thermodynamic limit: 
it is deterministic, with either weak or strong damping, 
temperature increases under compression; the evolution is independent 
of $M$  (for $M$ large enough) and it proceeds until mechanical 
equilibrium   $p^+=p^-$ is reached. In the second stage, 
which is now characterised by a time scale of order
 $\tau_2 =M\tau_1/m$, the evolution has exactly 
the opposite properties: it is stochastic, which implies
 heat transfer through the piston, and the system evolves 
with constant pressure $i.e.$ $p^{\pm}(t)=p_f + {\mathcal{O}}
 (1/M)$, to a state of thermal equilibrium $(T^-=T^+)$; in this second stage the evolution depends strongly on $M$, more precisely we observe the scaling relation $X_M(t)= X(t/M)$
where $X(t)$ is independent of $N^{\pm}$ and $M$.
\vskip 15mm
\noindent
{\bf Acknowledgements:} The authors are grateful to J.L. Lebowitz, J. Piasecki
and Ya. Sinai for stimulating informations about their own work
on a similar model. Ch. Gruber is grateful to J.L. 
Lebowitz for his  suggestion to study the thermodynamic limit.
He would  like to thank  J.R. Dorfmann for his friendly invitation
where this work was started and the many discussions on Liouville
equations with movable boundary conditions. He is also grateful to Ya. Sinai
for his numerous e-mails on the problem, his warm hospitality in Princeton,
and his constructive criticisms, together with J.L. Lebowitz,
 concerning the Boltzmann
 approach previously considered.
A. Lesne greatly acknowledges the hospitality at the
Institute of Theoretical Physics of the \'Ecole Polytechnique F\'ed\'erale
de Lausanne where a part of this research has been performed.

\newpage
\section*{ Figure Captions}
\vskip 10mm
\noindent
{\bf Figure 1:} Thermodynamic limit. Position of the piston in function of time fixed for
$R=\bar{N}/ M=4$ and \\ a) $M=10\;000$, b) $M=25\;000$, c) $M=50\;000$, d) $M=100\;000$. Predicted mechanical  equilibrium for  adiabatic piston $X_{f}=8.42\;10^{4}$. Thermal equilibrium for conducting piston $X_{th }=30\;10^{4}$. 

\vskip 10mm
\noindent
{\bf Figure 2:}  From weak to strong damping. Position of the piston in function of time.
a) $R=0.1$ and $M=100\;000\;$,
b) $R=0.2$ and $M=100\;000\;$,
c) $R=4$ and $M=100\;000\;$,
d) $R=10$ and $M=30\;000\;$.

\vskip 10mm
\noindent
{\bf Figure 3:} Strong damping. Position of the piston in function of time.
a) $R=50$ and $M=6000\;$, b)$R=50$ and $M=10\;000$, c)$R=100$ and $M=4000\;$, d)$R=100$ and $M=6000$.

\vskip 10mm
\noindent
{\bf Figure 4:} Adiabatic evolution: approach to mechanical equilibrium
for $R=10$ and $M=30\;000$.
(a) Position.
(b) Pressure. (c) Temperature.
The predicted value for the mechanical  equilibrium are $X_{f}=8.42\;10^{4}$, $p^-_{f} = p^+_{f} = 5.5$, $T^-_{f}=1.54$, $T^+_{f}=9.46$.

\vskip 10mm
\noindent
{\bf Table 1:} From weak to strong damping. Simulations with $M$ sufficiently large to describe the thermodynamic limit. Value of the first minimum of the position of the piston and time at which it occurs. Comparison between the observed  and predicted period of oscillations.

\newpage
\begin{tabular}{c|cc|c|c|c}
&&&&&\\
\rule{10pt}{0pt}$R$ \rule{10pt}{0pt}& First&minimum & period of  & period of & period of \\
 & $X$ \rule{10pt}{0pt} &\rule{10pt}{0pt} $t$   &\rule{3pt}{0pt}
 oscillation
 \rule{3pt}{0pt} & \rule{3pt}{0pt}oscillation  \rule{3pt}{0pt} & 
\rule{3pt}{0pt}oscillation  \rule{3pt}{0pt}\\
& & & (simulation) & (predicted) & at $t=30\;.\;10^5$ 

(simulation)\\
&&&&&\\
\hline
&&&&&\\
0.1\hspace{10mm}& 6.6  $\;\times 10^{4}$ &3.7  $\;\times 10^{5}$ &7.34  $\;\times 10^{5}$ & 7.34 $\;\times 10^{5}$ &7.34 $\;\times 10^{5}$ \\
0.2&6.85&2.7&5.4&5.29&5.4\\
1&7.1&1.45&2.94&2.85&2.94\\
2&7.2&1.20&2.6&3.05& \\
4&7.4&1.0& $\;$& $\;$ &2.2\\
5&7.4&0.95&&&2.1\\
10&7.5&0.96&&&2.0\\
50&7.6&0.9&&& \\
100&7.6&0.9&&&\\
200&7.6&0.9&&&\\
300&7.6&0.9&&&\\
\end{tabular}
\vskip 10mm
\vskip 10mm
\noindent
{\bf Table 1}

\newpage
\hspace*{20mm}
\leavevmode
\epsfxsize= 90pt
\epsffile[10 15 200  700]{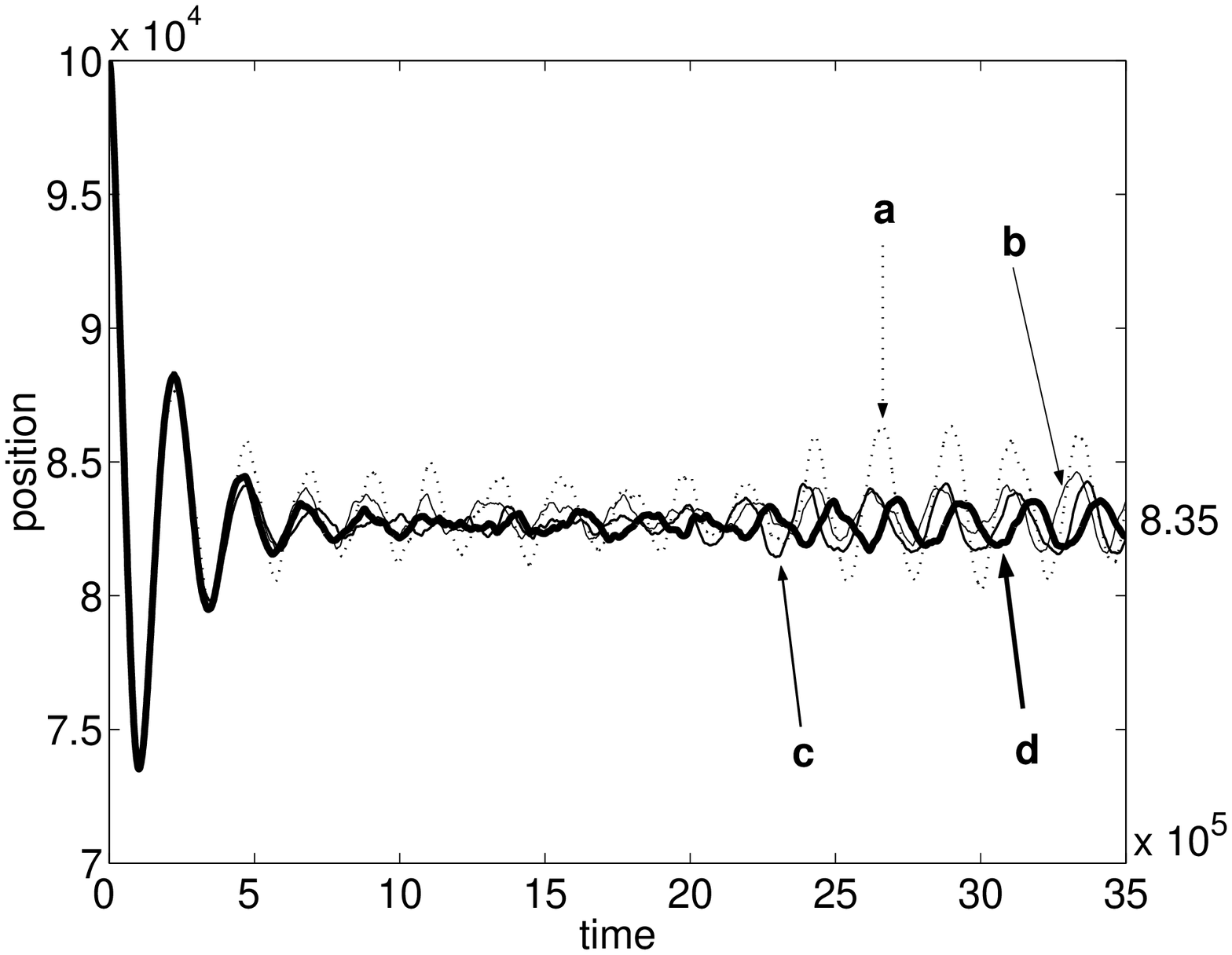}

{\bf Figure 1}
\newpage
\hspace*{20mm}
\leavevmode
\epsfxsize= 90pt
\epsffile[10 15 200  700]{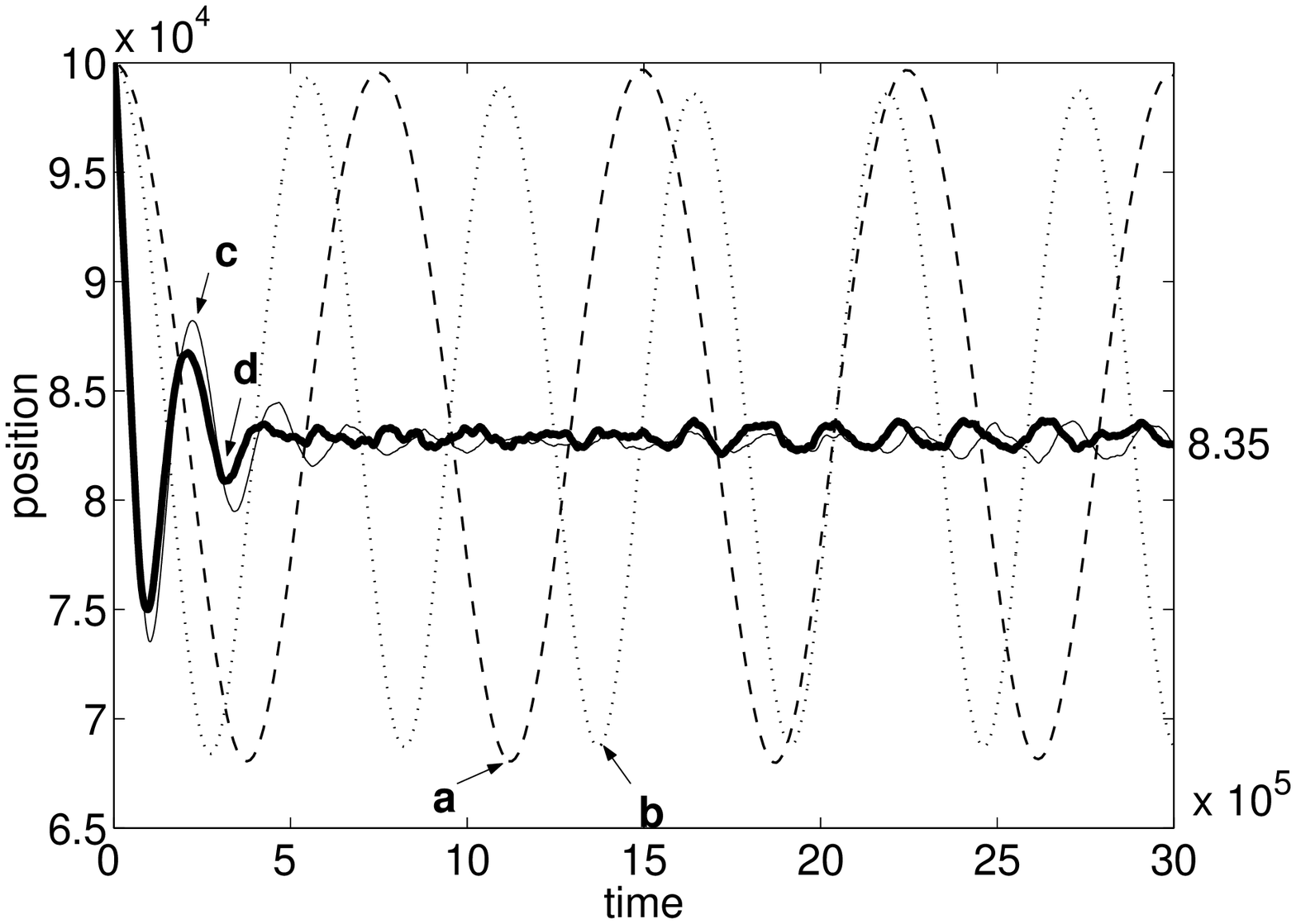}

{\bf Figure 2}
\newpage
\hspace*{20mm}
\leavevmode
\epsfxsize= 90pt
\epsffile[10 15 200  700]{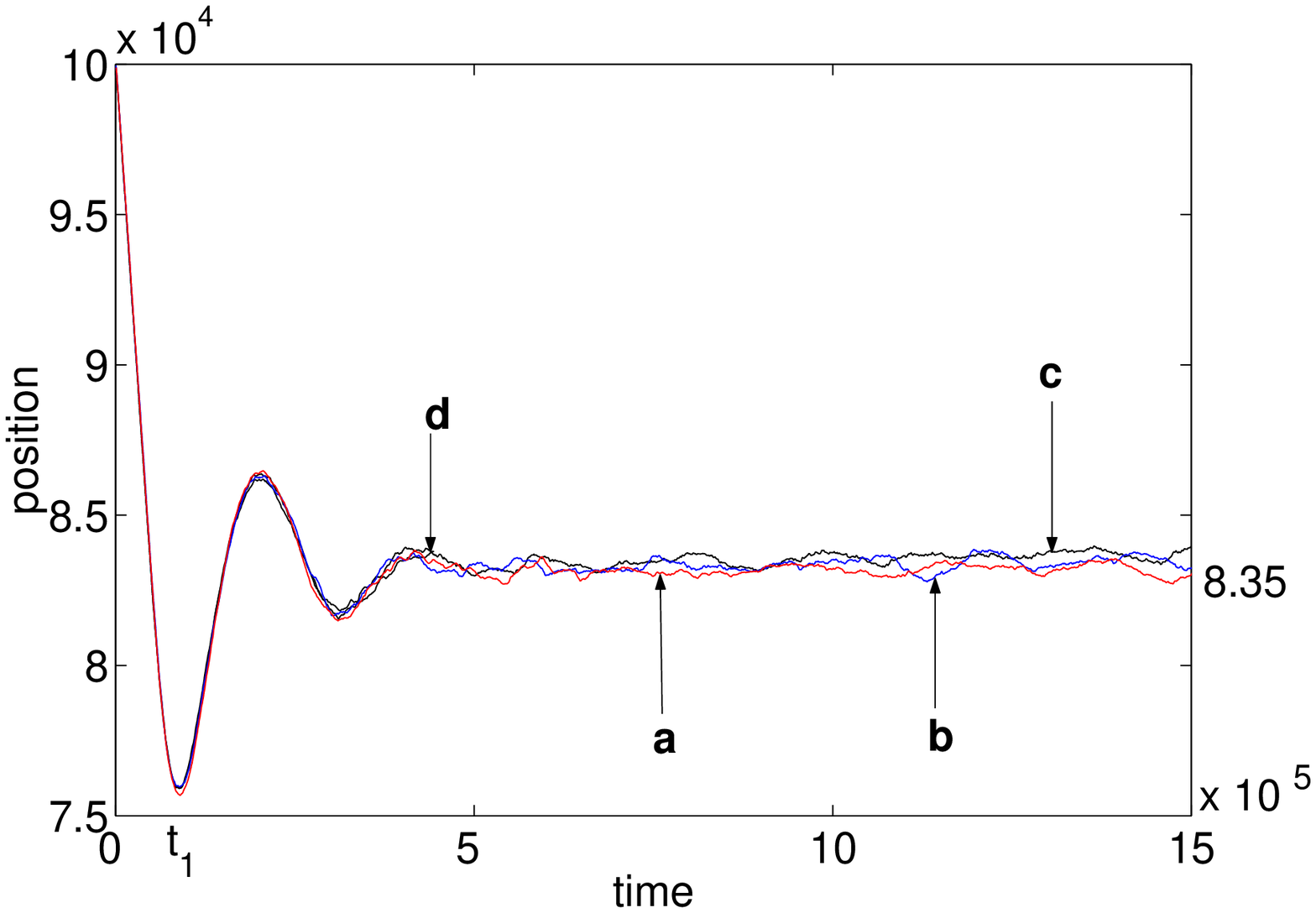}

{\bf Figure 3}

\newpage
\hspace*{50mm}
\leavevmode
\epsfxsize= 40pt
\epsffile[100 150 200  700]{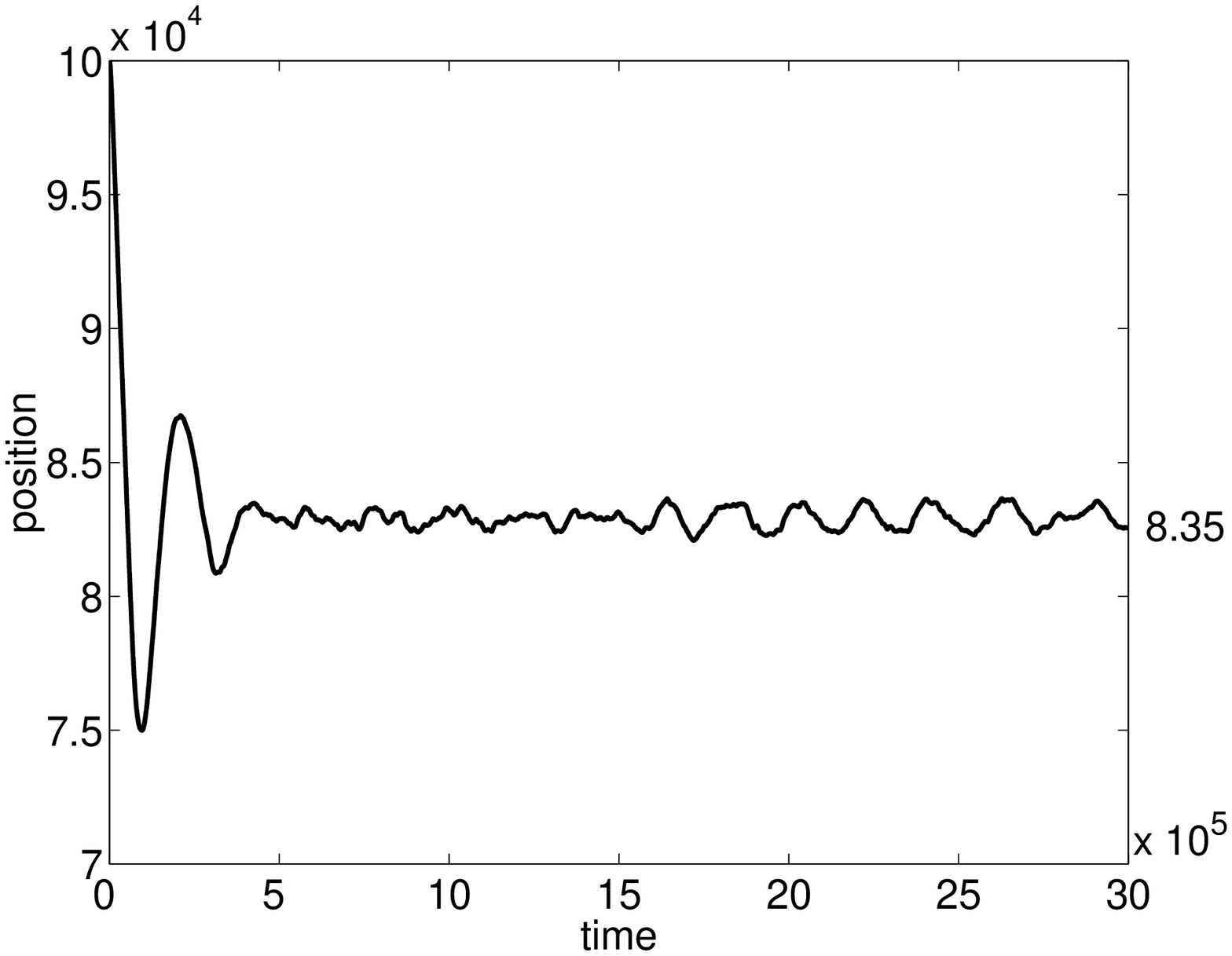}
\hspace*{15mm}(a)
\vspace*{-10mm}

\hspace*{50mm}
\epsfxsize= 40pt
\epsffile[100 150 200  700]{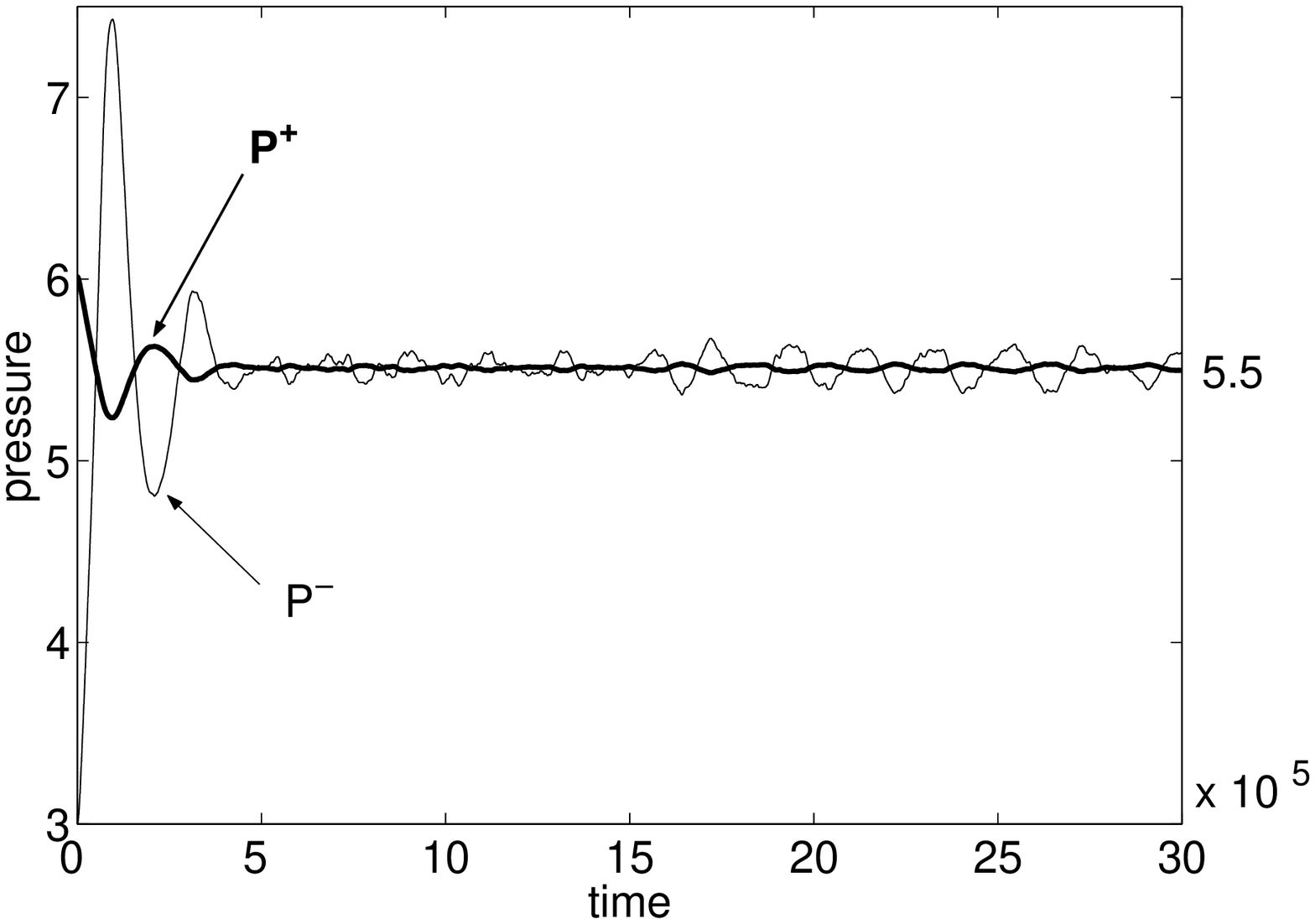}
\hspace*{15mm}(b)
\vspace*{-10mm}

\hspace*{50mm}
\epsfxsize= 40pt
\epsffile[100 150 200  700]{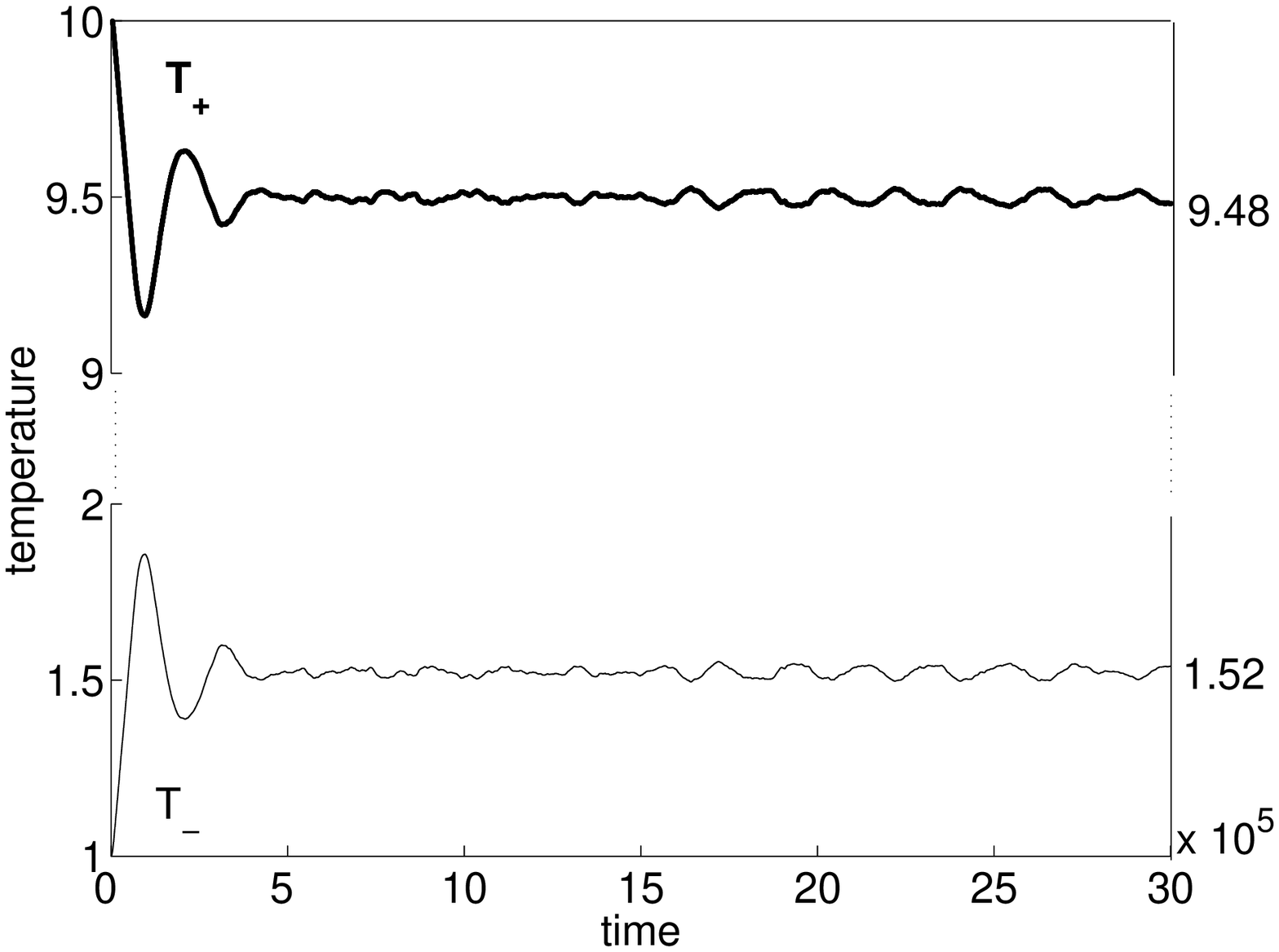}
\hspace*{15mm}(c)

\vspace*{10mm}

{\bf Figure 4}

\end{document}